\documentclass[aps,prb,superscriptaddress,showpacs]{revtex4-1}
\usepackage{graphicx,amsmath,amssymb,xspace,epsfig,float,multirow,subfigure,tabularx}

\pdfoutput=1

\begin{document}

\title{Effect of the type I to type II Weyl semimetal topological transition
on superconductivity}
\author{Dingping Li}
\email{lidp@pku.edu.cn}
\affiliation{School of Physics, Peking University, Beijing 100871, \textit{China}}
\affiliation{Collaborative Innovation Center of Quantum Matter, Beijing, China}
\author{Baruch Rosenstein}
\email{vortexbar@yahoo.com,correspondent author}
\affiliation{Electrophysics Department, National Chiao Tung University, Hsinchu 30050,
\textit{Taiwan, R. O. C}}
\author{B.Ya. Shapiro}
\email{shapib@mail.biu.ac.il}
\affiliation{Physics Department, Bar-Ilan University, 52900 Ramat-Gan, Israel}
\author{I. Shapiro}
\affiliation{Physics Department, Bar-Ilan University, 52900 Ramat-Gan, Israel}
\date{\today}

\begin{abstract}
The influence of recently discovered topological transition between type I
and type II Weyl semi-metals on superconductivity is considered. A set of
Gorkov equations for weak superconductivity in Weyl semi-metal under
topological phase transition is derived and solved. The critical temperature
and superconducting gap both have spike in the point the transition point as
function of the tilt parameter of the Dirac cone determined in turn by the
material parameters like pressure. The spectrum of superconducting
excitations is different in two phases: the sharp cone pinnacle is
characteristic for a type I, while two parallel almost flat bands, are
formed in type II. Spectral density is calculated on both sides of
transition demonstrate different weight of the bands. The superconductivity
thus can be used as a clear indicator for the topological transformation.
Results are discussed in the light of recent experiments.
\end{abstract}

\maketitle
\affiliation{School of Physics, Peking University, Beijing 100871, \textit{China}}
\affiliation{Collaborative Innovation Center of Quantum Matter, Beijing, China}
\affiliation{Electrophysics Department, National Chiao Tung University, Hsinchu 30050,
\textit{Taiwan, R. O. C}}
\affiliation{Physics Department, Bar-Ilan University, 52900 Ramat-Gan, Israel}
\affiliation{Physics Department, Bar-Ilan University, 52900 Ramat-Gan, Israel}

\section{Introduction}

Effect of the Fermi surface topology on properties of a crystalline
conductor was a subject of theoretical investigations over the years \cite%
{Lifshitsbook}. Experimentally a continuous deformation of the Fermi surface
(without changing the chemical nature of the crystal) can be experimentally
achieved by external factors like pressure and electric field. Historically
transitions associated with change of topology were called "$2.5$
transition" \cite{2.5}. Transport in materials with low electron density is
especially sensitive to modification of the Fermi surface \cite{Varlamov}
while these transitions were predicted to make a great impact on the
superconducting state \cite{2.5 super}. Recently new class of such materials
made the phenomenon important in a very different setting. A broad number of
low electron density 2D and 3D Weyl semi-metals were discovered. Most of
them are characterized by linear dispersion relation near the Fermi surface.
Topology and even dimensionality of the small Fermi surface in these
materials is linked to the Berry phases of their band structure\cite{Volovik}%
. The first material of this class, graphene\cite{Geim}, exhibits the
highest symmetry leading to linear ultra - relativistic spectrum, however
most of the other materials are anisotropic. Examples include 2D Weyl
semi-metals (WSM) silicene, germanene and borophene \cite{borophene} and 3D
crystals\cite{neupane} $Na_{3}Bi\ $, and\cite{wang1} $Cd_{3}As_{2}$ and
numerous layered organic compounds\cite{organic,Goerbig1,Goerbig2}.
Topological insulators (TI) like $Bi_{2}Se_{3}$ and other \cite{ARPES}
generally have Dirac cones on their surfaces of topological insulators.

It was realized recently that this variety of novel materials should be
differentiated between the more isotropic "type I" Weyl semimetals \cite%
{Soluyanov,Sun} see Fig.1a, and highly anisotropic type II Weyl semi-metals
in which the cone of the linear dispersion relation is tilted beyond a
critical angle (Fig.1b). The newer type-II Weyl fermion materials, $WTe_{2}$
exhibits exotic phenomena such as angle dependent chiral anomaly \cite%
{Soluyanov}. Discovery of a Weyl semimetal in $TaAs$ offers the first Weyl
fermion observed in nature and dramatically broadens the classification of
topological phases. Other dichalkogenides like \cite{MoTe} $MoTe_{2}$ and
\cite{PtTe} $PtTe_{2}$ were demonstrated to be type II Weyl semi - metals.
In particular, the series $Mo_{x}W_{1-x}Te_{2}$ inversion-breaking, layered,
tunable semimetals is already under study as a promising platform for new
electronics and recently proposed to host Type II Weyl fermions \cite{MoTeW}%
. Some other materials, like layered organic compound $\alpha
-(BEDT-TTF)_{2}I_{3}$, were long suspected \cite{Goerbig1} to be a 2D
type-II Dirac fermion. Several classes of materials were predicted via band
structure calculations to undergo the I to II transition while doping or
pressure is changed \cite{band}.

Theoretically physics of the transitions between the type I to type II Weyl
semi-metals were considered in the context of superfluid phase\cite{Volovik}
A of $He_{3}$, layered organic materials in 2D\cite{Goerbig2} and 3D Weyl
semi-metals \cite{Ye}. The pressure modifies the spin orbit coupling that in
turn determines the topology of the Fermi surface of these novel materials
\cite{Sun}. Very recently the topological transition in Weyl semi - metal
under pressure was observed \cite{toptransition}.

This transition is an ideal example of the "$2.5$ transition" mentioned
above. Both the Fermi surface reconstruction and low electron density are
present at the extreme in these materials. It is generally difficult to find
good indicators for such a transition. Since superconductivity is especially
affected by the "$2.5$ transition", it might serve as the indicator. Indeed
some Weyl materials are known to be superconducting. Surface of the well
known TI $Bi_{2}Se_{3}$ exhibits under certain conditions superconductivity
of up to\cite{Ong} $8K$, some organic materials like\cite{kappaI} $\kappa
-(BEDT-TTF)_{2}X$ with $X=I_{3}$ or other anion\cite{organicsuper} are
superconducting.

A detailed study of superconductivity in TI under hydrostatic pressure
revealed a curious dependence of critical temperature of the superconducting
transition on pressure \cite{MoTeSup} . In intercalated $%
Sr_{0.065}Bi_{2}Se_{3}$ single crystal \cite{SrBiSe} considered to be a pure
material ambient weak superconductivity first is suppressed, but at high
pressure of $6GPa$ reappears and reaches relatively high $T_{c}$ of $10K$
that persists till $80GPa$. The increase is not gradual, but rather abrupt
in the region of $15GPa$ \cite{SrBiSe}. Superconductivity in similar TI
compounds \cite{Maple,Kong},\cite{BiTe} $Bi_{2}Se_{3}$ , $Bi_{2}Te_{3}$ and
3D Weyl semi-metal\cite{HfTe} $HfTe_{5}$ were also studied experimentally.
The critical temperature $T_{c\text{ }}$ in some of these systems shows a
sharp maximum as a function of pressure. This contrasts with generally
smooth dependence on pressure in other superconductors (not suspected to be
Weyl materials) like a high $T_{c}$ cuprate\cite{YBCO} $YBCO$. Various
mechanisms of superconductivity in Dirac semi -- metals and topological
insulators turned superconductors have been considered theoretically \cite%
{DasSarma,FuBerg,frontiers}. A theory predicted possibility of
superconductivity in the type II Weyl semimetals was developed recently in
the framework of Eliashberg model \cite{Zyuzin}.

We show in this paper that superconducting critical temperature and energy
gap features is an efficient marker of the topological type I to type II
transition under pressure. The $s$-wave pairing in a general Weyl semi-metal
with tilted Dirac cones is considered in the framework of the boson mediated
adiabatic regime. We calculate the critical temperature $T_{c}$ and the
energy gap $\Delta $ at zero temperature for arbitrary tilt (controlled by
pressure in certain materials) and find their sharp increase at topological
transition point from type I to type II Weyl semi-metal.

\section{Type I and type II Weyl semi - metal with local pairing interaction.%
}

Weyl material typically possesses several sublattices. We exemplify the
effect of the topological transition on superconductivity using the simplest
possible model with just two sublattices denoted by $\alpha =1,2$. The band
structure near the Fermi level of a 2D Weyl semi-metal is well captured by
the following Hamiltonian,
\begin{eqnarray}
K &=&\int_{\mathbf{r}}\psi _{\alpha }^{sL+}\left( \mathbf{r}\right)
K_{\alpha \beta }^{L}\psi _{\beta }^{sL}\left( \mathbf{r}\right) +\psi
_{\alpha }^{sR+}\left( \mathbf{r}\right) K_{\alpha \beta }^{R}\psi _{\beta
}^{sR}\left( \mathbf{r}\right) \text{;\ \ \ \ }  \label{kinetic} \\
\text{\ }K_{\alpha \beta }^{L,R} &=&-i\hbar v\left( \nabla _{x}\sigma
_{\alpha \beta }^{x}\mp \nabla _{y}\sigma _{\alpha \beta }^{y}\right)
+\left( -i\hbar \mathbf{w}\cdot \mathbf{\nabla }-\mu \right) \delta _{\gamma
\delta }\text{.}  \notag
\end{eqnarray}%
Here $v$ is Fermi velocity, $\mu $ - chemical potential, $\sigma $ are Pauli
matrices in the sublattice space and $s$ is spin projection. Generally there
are a number of pairs of points (Dirac cones) constituting the Fermi
"surface" of such a material at chemical potential $\mu =0$. We restrict
ourself to the case of just one left handed ($L$) and one right handed ($R$)
Dirac points, typically but not always separated in the Brillouin zone.
Generalization to several pairs is straightforward. The 2D velocity vector $%
\mathbf{w}$ defines the tilt of the cone, see dispersion relation of one of
the cones in Fig.1. The graphene - like dispersion relation in Fig.1a for $%
w=0.5$ represents the type I Weyl semi-metal, while when the length of the
tilt vector exceeds $v$, Fig. 1b, the material becomes a type II Weyl semi -
metal.
\begin{figure}[tbp]
\centering
\includegraphics[width=16cm]{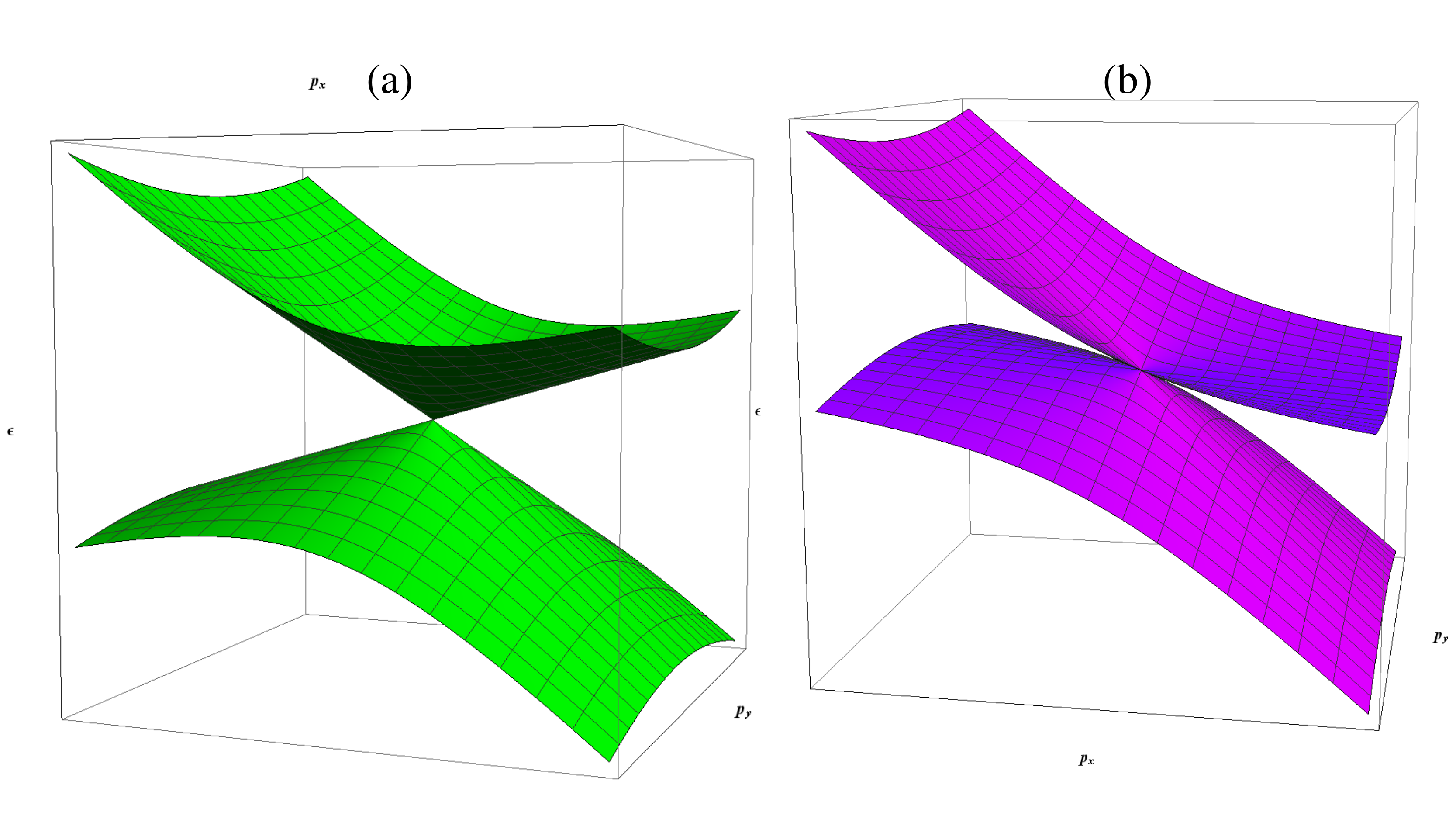}
\caption{ Spectrum of normal Weyl semimetal a. type I $\left( w/v=0.5\right)
$. b. Strongly tilted Dirac cone for the type II semimetals $\left(
w/v=1.2\right) $. }
\end{figure}

The effective electron-electron attraction due to either electron - phonon
attraction and Coulomb repulsion (pseudopotential) or some unconventional
pairing mechanism creates pairing. We assume that different valleys are
paired independently and drop the valley indices (multiplying the density of
states by $2N_{f}$). Further we assume the local singlet $s$-channel
interaction Hamiltonian
\begin{equation}
V=\frac{g^{2}}{2}\ \int d\mathbf{r}\text{ }\psi _{\alpha }^{+\uparrow
}\left( \mathbf{r}\right) \psi _{\beta }^{\downarrow +}\left( \mathbf{r}%
\right) \psi _{\beta }^{\uparrow }\left( \mathbf{r}\right) \psi _{\alpha
}^{\downarrow }\left( \mathbf{r}\right) \text{.}  \label{int}
\end{equation}%
As usual the interaction has a cutoff frequency $\Omega $, so that it is
active in an energy shell of width $2\hbar \Omega $ around the Fermi level
\cite{Abrikosov}. For the phonon mechanism it is the Debye frequency.

\section{Spectrum of excitations in the superconducting state.}

Finite temperature properties of the condensate are described by the normal
and the anomalous Matsubara Greens functions \cite{Abrikosov},%
\begin{eqnarray}
G_{\alpha \beta }^{ts}\left( \mathbf{r}\tau ,\mathbf{r}^{\prime }\tau
^{\prime }\right) &=&-\left\langle T_{\tau }\psi _{\alpha }^{t}\left(
\mathbf{r}\tau \right) \psi _{\beta }^{s+}\left( \mathbf{r}^{\prime }\tau
^{\prime }\right) \right\rangle =\delta ^{ts}g_{\alpha \beta }\left( \mathbf{%
r-r}^{\prime },\tau -\tau ^{\prime }\right) ;  \label{green} \\
F_{\alpha \beta }^{ts}\left( \mathbf{r}\tau ,\mathbf{r}^{\prime }\tau
^{\prime }\right) &=&\left\langle T_{\tau }\psi _{\alpha }^{t}\left( \mathbf{%
r}\tau \right) \psi _{\beta }^{s}\left( \mathbf{r}^{\prime }\tau ^{\prime
}\right) \right\rangle =-\varepsilon ^{ts}f_{\alpha \beta }\left( \mathbf{r-r%
}^{\prime },\tau -\tau ^{\prime }\right) ;  \notag \\
F_{\alpha \beta }^{+ts}\left( \mathbf{r}\tau ,\mathbf{r}^{\prime }\tau
^{\prime }\right) &=&\left\langle T_{\tau }\psi _{\alpha }^{t+}\left(
\mathbf{r}\tau \right) \psi _{\beta }^{s+}\left( \mathbf{r}^{\prime }\tau
^{\prime }\right) \right\rangle =\varepsilon ^{ts}f_{\alpha \beta
}^{+}\left( \mathbf{r-r}^{\prime },\tau -\tau ^{\prime }\right) .  \notag
\end{eqnarray}%
The second equality in each line relies on homogeneity and unbroken
invariance under spin rotations. The gap function in the s-wave channel is%
\begin{equation}
\Delta _{\alpha \gamma }\equiv -\frac{g^{2}}{4}\varepsilon
^{s_{3}s_{2}}\left\langle \psi _{\alpha }^{s_{3}}\left( \mathbf{r,}\tau
\right) \psi _{\gamma }^{s_{2}}\left( \mathbf{r,}\tau \right) \right\rangle
=\sigma _{\alpha \gamma }^{x}\Delta .  \label{n5}
\end{equation}%
In terms of Fourier transforms, $g_{\gamma \kappa }\left( r,\tau \right)
=T\sum\nolimits_{\omega \mathbf{p}}\exp \left[ i\left( -\omega \tau +\mathbf{%
p\cdot r}\right) \right] g_{\gamma \kappa }\left( \omega ,p\right) $, the
Gorkov equations read (see Appendix A):
\begin{eqnarray}
\left( v\mathbf{p}\cdot \mathbf{\sigma }_{\gamma \beta }+\left( i\omega +\mu
-wp_{x}\right) \delta _{\gamma \beta }\right) g_{\beta \kappa }\left( \omega
,p\right) +\Delta \sigma _{\alpha \gamma }^{x}f_{\alpha \kappa }^{+}\left(
\omega ,p\right) &=&\delta ^{\gamma \kappa };  \label{Gorkoveq} \\
\left( v\mathbf{p}\cdot \mathbf{\sigma }_{\beta \gamma }+\left( -i\omega
+\mu -wp_{x}\right) \delta _{\gamma \beta }\right) f_{\beta \kappa
}^{+}\left( \omega ,p\right) -\Delta ^{\ast }\sigma _{\gamma \alpha
}^{x}g_{\alpha \kappa }\left( \omega ,p\right) &=&0\text{.}  \notag
\end{eqnarray}%
The Matsubara frequency at temperature $T$ takes values $\omega _{n}=\pi
T\left( 2n+1\right) $, and units are chosen such that $\hbar =1$, We have
chosen coordinates in such a way that the vector $w$ causing the tilt of the
Dirac cone is oriented along the $x$ axis.
\begin{figure}[tbp]
\centering
\includegraphics[width=16cm]{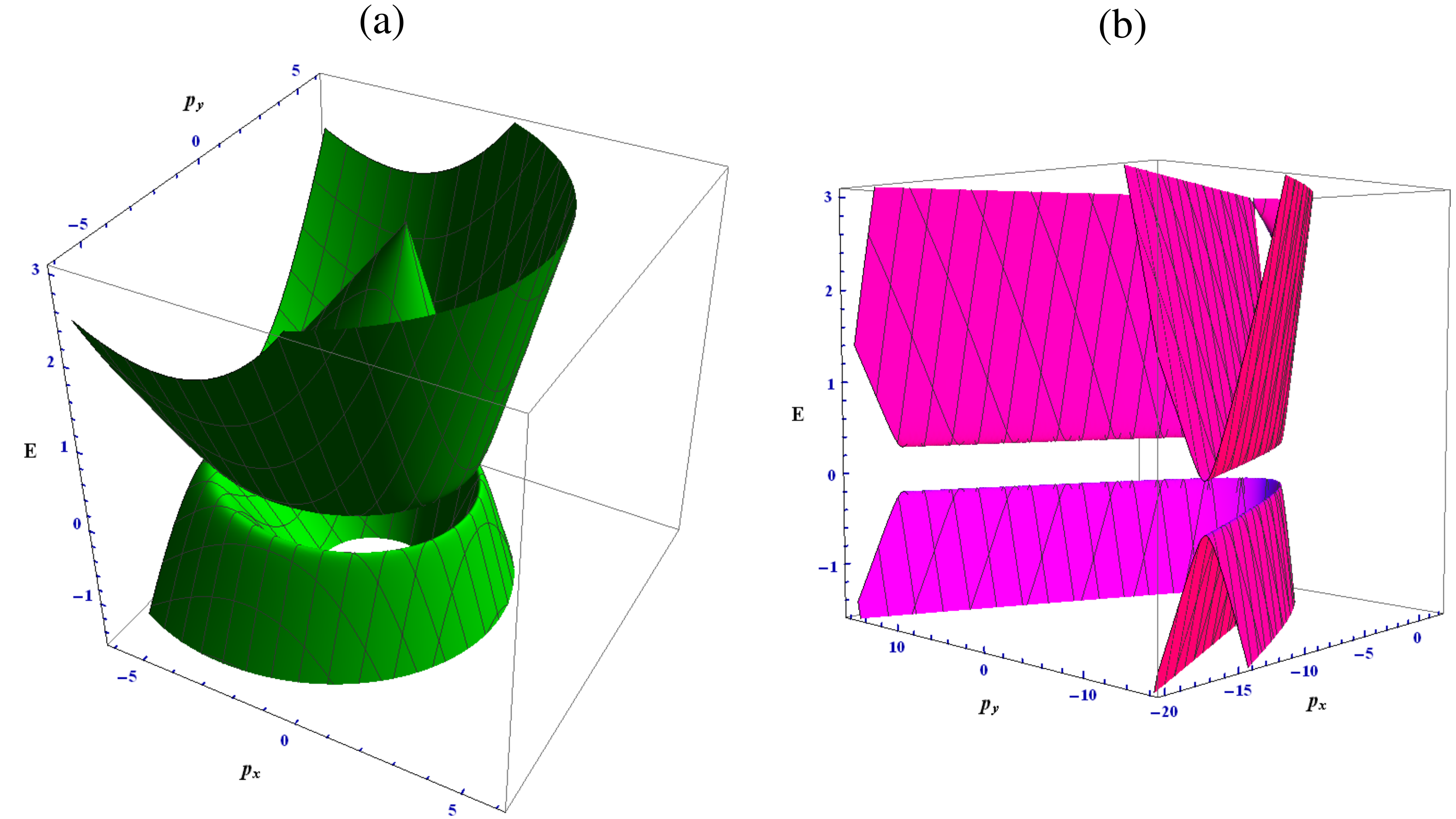}
\caption{Spectrum of superconducting quasiparticles as function of momentum
(in units of $\hbar \Omega /v$) for (a) Type I ($w/v=0.5$) and (b) Type II ($%
w/v=1.2$) Weyl semimetal for chemical potential $\protect\mu =3$ and energy
gap $\Delta =0.25$ (in units of phonon energy $\hbar \Omega $) }
\label{Fig.2}
\end{figure}
In matrix form (in sub-lattice space) we obtain the solution (see Appendix
A)
\begin{eqnarray}
\widehat{g}\left( \omega ,\mathbf{p}\right) &=&\sigma ^{x}\left( \mathbf{p}%
\cdot \mathbf{\sigma }^{t}+\left( -i\omega +\mu -wp_{x}\right) I\right)
M^{-1};  \label{solution} \\
\widehat{f}^{+}\left( \omega ,\mathbf{p}\right) &=&M^{-1}\Delta ^{\ast }%
\text{,}  \notag
\end{eqnarray}%
where
\begin{equation}
M=\left( v\mathbf{p}\cdot \mathbf{\sigma }+\left( i\omega +\mu
-wp_{x}\right) I\right) \sigma ^{x}\left( v\mathbf{p}\cdot \mathbf{\sigma }%
^{t}+\left( -i\omega +\mu -wp_{x}\right) I\right) +\left\vert \Delta
\right\vert ^{2}\sigma ^{x}\text{,}  \label{delt2}
\end{equation}%
and $I$ is the identity matrix. The determinant of $M$,%
\begin{equation}
\det M=-\left( \Delta ^{2}+\omega ^{2}+\left( vp-\mu +wp_{x}\right)
^{2}\right) \left( \Delta ^{2}+\omega ^{2}+\left( -vp-\mu +wp_{x}\right)
^{2}\right)  \label{det}
\end{equation}%
continued to physical energy, $\omega \rightarrow -iE$, gives the spectrum
of quasiparticles
\begin{equation}
E_{\pm }^{2}=\Delta ^{2}+\left( \pm vp-\mu +wp_{x}\right) ^{2}\text{,}
\label{delt3}
\end{equation}%
depicted for type I and type II Weyl semimetals in Fig.2.

The quasi-particle spectrum is very different. In a superconducting graphene
- like material for $\mu >>\Delta $, the energy gap is minimal on the circle
of radius $\mu /v$ , see Fig.2a. Note that the sharp cone pinnacle in the
excitations spectrum wedge is formed at the former cone location. For the
type II Weyl semi-metal turned superconductor, Fig.2b, the low energy
spectrum consists of two parallel flat bands,\ while the pinnacle disappears.

\subsection{Spectral density.}

The spectral density function
\begin{eqnarray}
A\left( E,\mathbf{p}\right) &=&-\frac{1}{\pi }Im\text{Tr}\ \widehat{g}\left(
-iE+\eta ,\mathbf{p}\right) ;  \label{spectral} \\
\text{Tr}\ \widehat{g}\left( \omega ,\mathbf{p}\right) &=&\frac{2}{\det M}%
\left\{ \left( v^{2}p^{2}-\left( \mu -wp_{x}-i\omega \right) ^{2}\right)
\left( \mu -wp_{x}+i\omega \right) -\Delta ^{2}\left( \mu -wp_{x}-i\omega
\right) \right\}  \notag
\end{eqnarray}%
presented in Fig. 3, where different weight of the dispersion law branches
accompanied Type I to Type II transition. Here $\eta $ is the "disorder"
parameter, $\widehat{g}\left( \omega ,\mathbf{p}\right) $ is defined by the
Eq.\ref{solution}).

\begin{figure}[tbp]
\centering
\includegraphics[width=16cm]{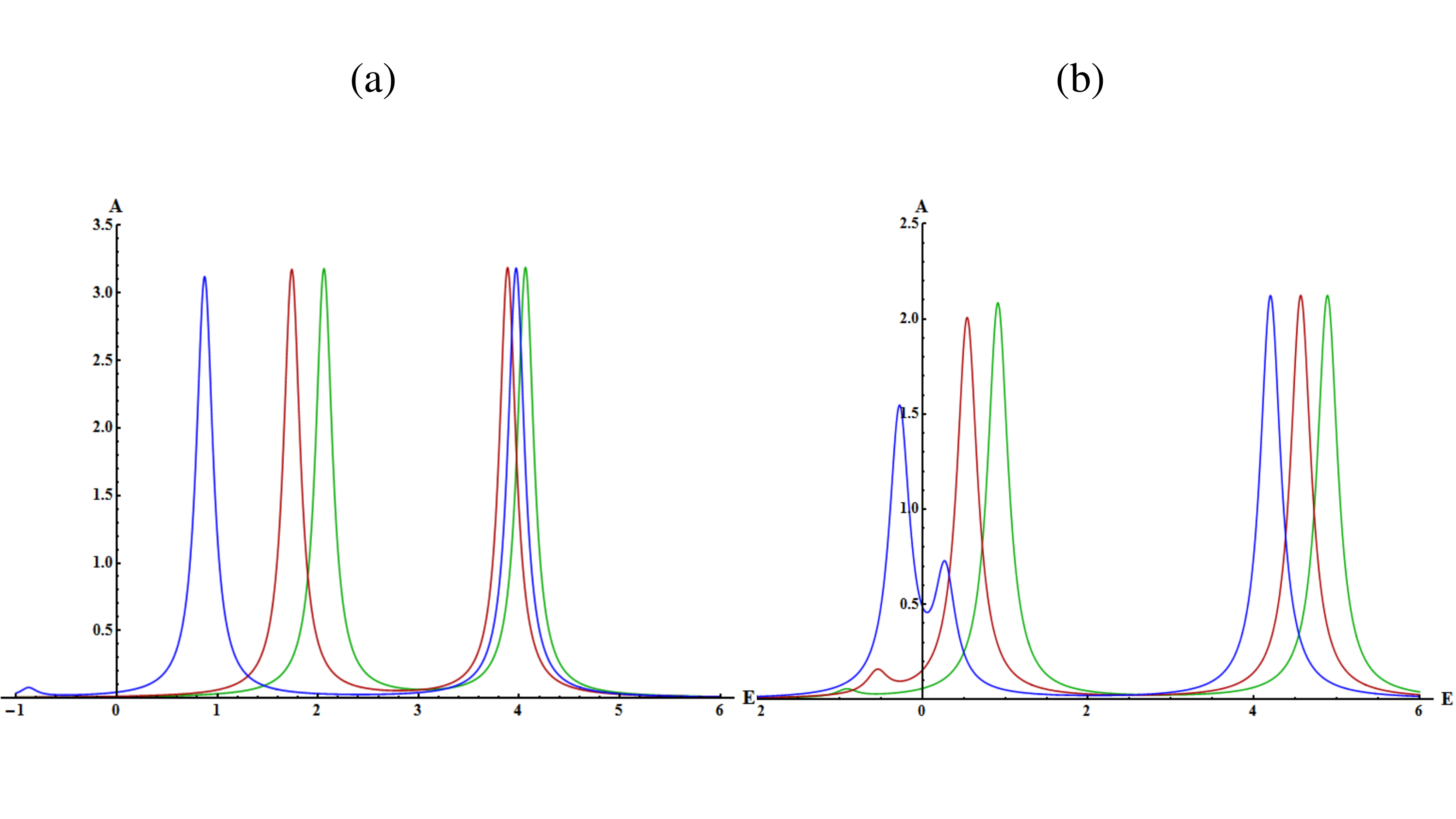}
\caption{Spectral density $A\left( \mathbf{p},E\right) $ for type I (a) $%
\left( w/v=0.5\right) $and type II (b) $\left( w/v=1.2\right) $
superconducting semimetals. Here $p_{x}=1,p_{y}=0.1$ (green), $0.4$ (red), $%
0.8$ (blue) in units of $\hbar \Omega /v$. Here $\protect\mu =3,$ $\Delta
=0.25$ (in the units of cut off phonon energy $\hbar \Omega ).$}
\end{figure}

\section{Critical temperature and the energy gap.}

The gap as function of the material parameters is determined by the equation
\begin{equation}
\Delta ^{\ast }=\frac{\Delta ^{\ast }}{2\left( 2\pi \right) ^{3}}%
T\sum\limits_{\omega }\int d\mathbf{p}\text{ Tr}\left[ \sigma _{x}M^{-1}%
\right] \text{,}  \label{GE}
\end{equation}%
resulting in at zero temperature
\begin{equation}
\frac{1}{g^{2}}=\frac{1}{\left( 2\pi \right) ^{3}}\int d\omega d\mathbf{p}%
\frac{v^{2}p^{2}+\Delta ^{2}+\left( \mu -wp_{x}\right) ^{2}+\omega ^{2}}{%
\left( \Delta ^{2}+\left( vp+wp_{x}-\mu \right) ^{2}+\omega ^{2}\right)
\left( \Delta ^{2}+\left( vp-wp_{x}+\mu \right) ^{2}+\omega ^{2}\right) }%
\text{.}  \label{gap2}
\end{equation}%
Performing integration on $\omega $ and momenta subject to restriction of
being inside the energy shell of $\Omega $ around the Fermi level, see Fig. 4

\begin{figure}[tbp]
\centering
\includegraphics[width=16cm]{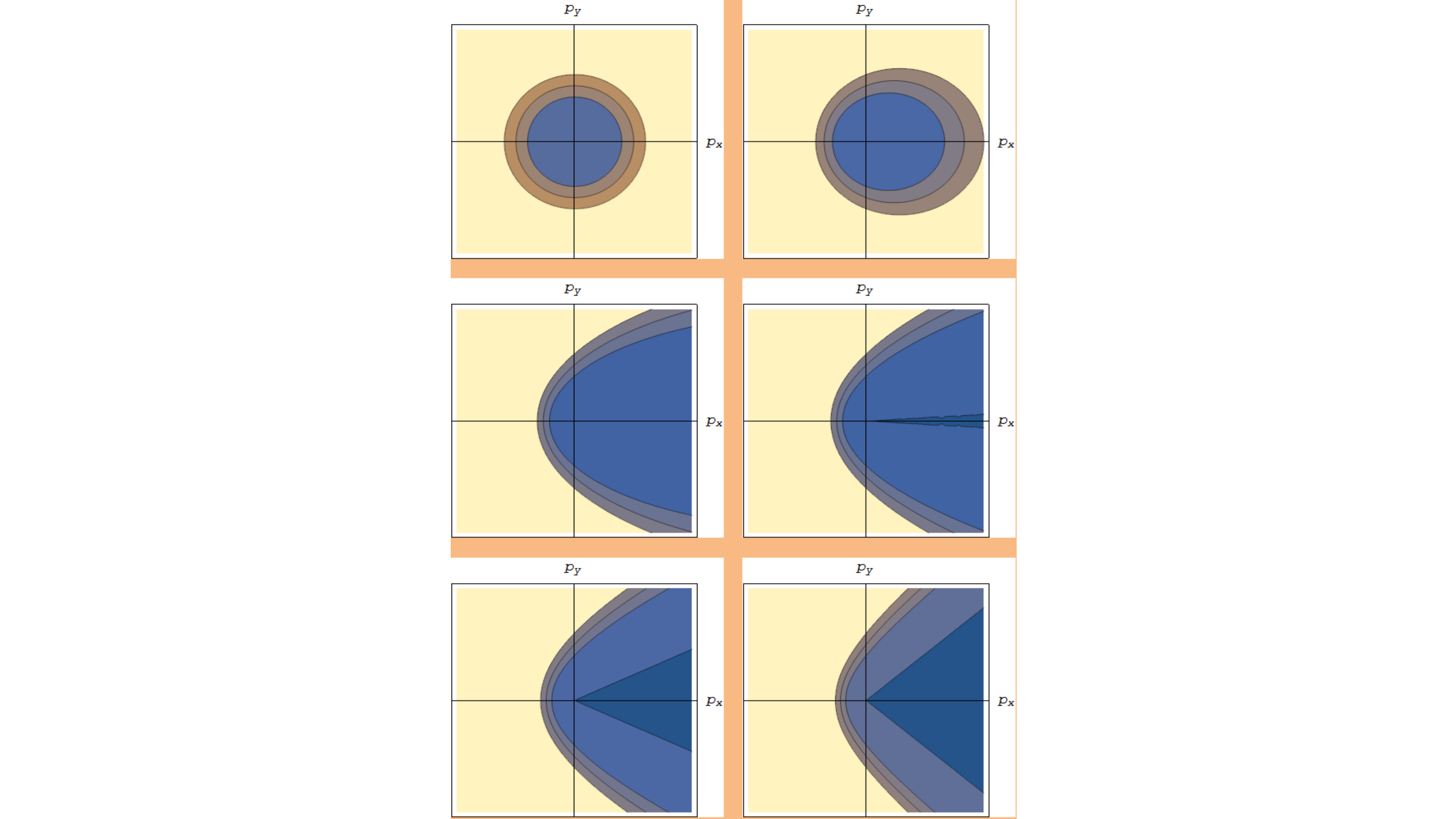}
\caption{Equipotential surfaces of the electron energy for different ratio $%
w/v$ (pairs of the figures from top to bottom): top figures line $w/v=0.01$%
(left), $w/v=0.4$ middle line: $w/v=0.91$ (left), $w/v=1.004$ and bottom
line: $w/v=1.1$ (left), $w/v=1.4.$}
\label{Fig4}
\end{figure}

one obtains in the adiabatic limit $\mu >>\Omega >>\Delta $, after
integration over azimuthal angle (see Appendix B for details),

\begin{equation}
\frac{1}{g^{2}}=\frac{\mu }{4\pi v^{2}}f\left( \kappa \right) \log \left[
\frac{2\Omega }{\Delta }\right] \text{.}  \label{gapeq1}
\end{equation}%
Here $\kappa =w/v$ is the anisotropy parameter. The function $f$ is:%
\begin{equation}
f\left( \kappa \right) =\left\{
\begin{array}{c}
\frac{2\ }{\ \left( 1-\kappa ^{2}\right) ^{3/2}}\ \text{ \ \ \ \ \ \ \ \ \ \
\ \ \ \ \ \ \ \ \ \ \ for }\kappa <1 \\
\frac{\ 2\kappa ^{2}}{\pi \left( \kappa ^{2}-1\right) ^{3/2}}\left\{ 2\sqrt{%
1+\kappa }-1+\log \frac{2\left( \kappa ^{2}-1\right) }{\kappa \left( 1+\sqrt{%
1+\kappa }\right) ^{2}\delta }\right\} \text{\ \ \ \ \ \ for }\kappa >1%
\end{array}%
\right. \text{,}  \label{f}
\end{equation}%
where $\delta =\pi a\Omega /v\hbar $ is the cutoff parameter, $a$ is the
interatomic distance. (The effective density of states $f\left( \kappa
\right) $ formally diverges at $\kappa =1.$ It's result of linear dispersion
relation in the model Weyl Hamiltonian. Indeed, \ the linear approximation
of the dispersion relation in the Weyl semimetal is valid only for small
neighborhood of the Dirac point. At higher energies the band spectrum is
nonlinear and cut off by the band width $1/a$. Actual cutoff is not very
important since the singularity is logarithmic).

The superconducting gap in this case therefore is (returning to physical
units)%
\begin{equation}
\Delta =2\hbar \Omega \exp \left[ -\frac{1}{\lambda f\left( \kappa \right) }%
\right] \text{.}  \label{gap}
\end{equation}%
where $\lambda =N\left( \mu \right) g^{2}$ is the conventional effective
electron - electron dimensionless coupling constant and $\log \left[ \gamma
_{E}\right] =0.577$. Dependence is presented in Fig.5. The effective density
of states (DOS) $D\left( E,\kappa \right) =N\left( E\rightarrow \mu \right)
f\left( \kappa \right) $ in the normal state for $N_{f}$ pairs of Dirac
cones is $\frac{2N_{f}}{4\pi v^{2}}\mu f\left( \kappa \right) $. This DOS
coincides with the spectral density given in Eq. \ref{spectral}) for $\Delta
=0$ integrated over momenta, $D\left( E,\kappa \right) =\int d\mathbf{p}$ $%
A\left( E,\mathbf{p},\kappa \right) $.

\begin{figure}[tbp]
\centering
\includegraphics[width=12cm]{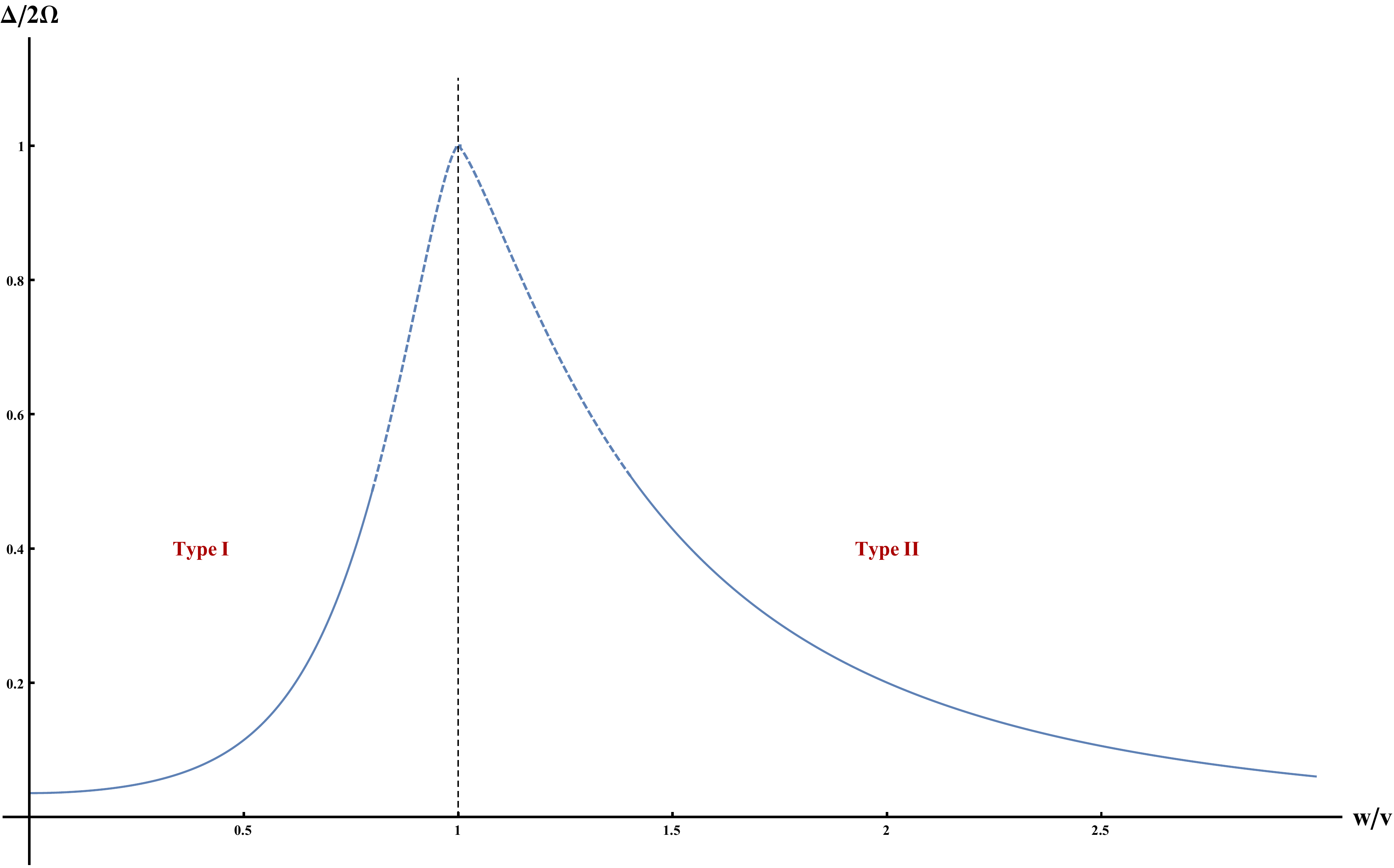}
\caption{Superconducting gap at zero temperature versus cone tilted
parameter $\protect\kappa $. (Dash line marks region beyond assumption $%
\protect\mu >>\Omega >>T_{c}).$ Cut off parameter $\protect\delta =0.01,%
\protect\mu =3,\protect\lambda =0.15.$}
\label{Fig5}
\end{figure}
The critical temperature is defined from the gap equation, when the gap $%
\Delta $ vanishes:

\begin{equation}
\frac{1}{g^{2}}=\frac{T}{\left( 2\pi \right) ^{2}}\int d\mathbf{p}%
\sum\nolimits_{n}\frac{v^{2}p^{2}+\left( \mu -wp_{x}\right) ^{2}+\omega
_{n}^{2}}{\left[ \left( vp+wp_{x}-\mu \right) ^{2}+\omega _{n}^{2}\right] %
\left[ \left( vp-wp_{x}+\mu \right) ^{2}+\omega _{n}^{2}\right] }\text{.}
\label{C1}
\end{equation}%
The sum over Matsubara frequencies can be done,$\ \ $%
\begin{equation}
\frac{1}{g^{2}}=\frac{1}{4\left( 2\pi \right) ^{2}}\int d\mathbf{p}\left\{
\frac{\tanh \left[ \left\vert vp+wp_{x}-\mu \right\vert /2T\right] }{%
\left\vert vp+wp_{x}-\mu \right\vert }+\frac{\tanh \left[ \left\vert
vp+wp_{x}+\mu \right\vert /2T\right] }{\left\vert vp+wp_{x}+\mu \right\vert }%
\right\} .  \label{C2}
\end{equation}%
In the adiabatic approximation, $\mu >>\Omega $, performing integration one
obtains:

\begin{equation}
\frac{1}{g^{2}}\approx \frac{\mu f\left( \kappa \right) }{4\pi v^{2}}\left(
\log \frac{\Omega }{2T_{c}}\text{ }\tanh \left[ \frac{\Omega }{2T_{c}}\right]
-\int_{E=0}^{\Omega }\frac{\log E}{\cosh ^{2}\left[ E/2T_{c}\right] }\right)
\equiv f\left( \kappa \right) R\left( T_{c},\Omega ,\mu \right) \text{.}
\label{ct4}
\end{equation}

Eq.(\ref{ct4}) gives for critical temperature

\begin{equation}
T_{c}=1.14\Omega \exp \left[ -\frac{1}{\lambda f\left( \kappa \right) }%
\right] \text{,}  \label{tc1}
\end{equation}%
Fig.5 demonstrates that the critical temperature has a sharp spike at the
transition point $w=v$. The ratio $\frac{2\Delta }{T_{c}}$ is still
universal for any $w/v.$

To summarize, superconductivity in Weyl semi-metal with tilted Dirac cones
that in the normal phase undergoes a topological ("2.5") transition from
type I to type II semi-metal was theoretically considered. We calculated in
the framework of the phonon mediated pairing model the spectrum of the
superconducting quasi-particles, spectral density, critical temperature and
the superconducting gap as function of the tilt parameter $w/v$ .\ The
critical temperature and the superconducting energy gap of Weyl semi-metal
have a spike at the transition point $w/v=1$. The quasi-particle spectrum in
the superconducting state qualitatively discriminates between type I and
type II see Fig.2. In particular the sharp cone pinnacle in the excitations
spectrum wedge typical for a type I disappears, two parallel nearly flat
bands, are formed. The spectral density $A\left( \mathbf{p},E\right) $ is
also undergoing modification under type I to type II topological phase
transition, see Fig.3.

\section{Discussion and conclusions.}

We discuss next an application of the results to recent measurement on
superconductivity of a compound that have Weyl semi-metal signatures under
pressure.\ In particular, in intercalated $Sr_{0.065}Bi_{2}Se_{3}$ single
crystal superconductivity first is suppressed, but at high pressure of $6GPa$
reappears and reaches relatively high $T_{c}$ of $10K$ that persists till $%
80GPa$. The increase is not gradual, but rather abrupt in the region of $%
15GPa$.\ It was discovered recently that, while parameters of the system
(the electron - phonon coupling, Debye frequency $\Omega $) between two
structural phase transitions smoothly depend on a pressure, the tilt
parameter is very sensitive to a stress\cite{Sun,toptransition}. Our result
might provide an explanation of a spike in dependence of $T_{c}$ on pressure
in single crystals observed in many superconducting semi-metals \cite%
{SrBiSe,Maple,Kong}.

The experimental situation is not unambiguous. There are structural phase
transitions that might cause an abrupt change of electron - phonon coupling.
Efforts have been made to associate the changes of $T_{c}$ with such
transitions. For example while in \cite{Maple} such maximum in $T_{c}$ takes
place close to the structural phase transition from $C2/m$ (7-fold) to\
phase $C2/m$ (bcc). On the other hand results presented in \cite{Kong}
\textit{are clearly separated from structural phase transitions giving an
evidence that this maximum is caused by undergoing Type I to Type II
topological transition in semimetals}.

Problem of the applicability of the BCS approach and a related issue of
Migdal's theorem i.e. adiabatic limit $\left( \mu >>\Omega \right) $ in Weyl
semi-metals has been addressed in detail by DasSarma's group and in our
earlier paper \cite{DasSarma}.The energy gap in this systems does not exceed
ten Kelvins, while $\Omega $ is a hundred of Kelvins and chemical potential
in thousands as in conventional optical phonon BCS. This justifies our BCS
approach.

Although a very specific simple model of two dimensional single pair of
Dirac cones was employed, it is expected that generalization to similar
physical systems like the surfaces of 3D topological insulators and 3D Weyl
semi - metals will lead to similar conclusions.

\textit{Acknowledgements.}

We are grateful to T. Maniv, T. W. Luo, M. Lewkowicz, C. C. Hou, E. Farber
for valuable discussions. B.R. was supported by NSC of R.O.C. Grants No.
103-2112-M-009-014-MY3 and MOE ATU program. The work of D.L. also is
supported by National Natural Science Foundation of China (No. 11274018 and
No. 11674007). B.R. is grateful to School of Physics of Peking University
and Bar Ilan Center for Superconductivity for hospitality.

\section{Appendix A: Gorkov equations for a two sublattice system.}

In this Appendix Gorkov equations for a two sublattice system (index $\alpha
=1,2$) in Matsubara representation $t=-i\tau $ are briefly derived.

Equation of motion for creation and annihilation Matsubara operators read $%
\left( \hbar =1\right) $%
\begin{equation}
\frac{\partial \psi _{\gamma }^{+}}{\partial \tau }=\left[ H,\psi _{\gamma
}^{+}\right] ;\frac{\partial \psi _{\gamma }}{\partial \tau }=\left[ H,\psi
_{\gamma }\right] .  \label{A1}
\end{equation}%
Defining the Matsubara Green function as%
\begin{eqnarray}
G_{\alpha \beta }^{ts}\left( X,X^{\prime }\right) &=&\delta ^{ts}g_{\alpha
\beta }\left( X,X^{\prime }\right) =-\left\langle T_{\tau }\left\{ \psi
_{\alpha }^{t}\left( X\right) \psi _{\beta }^{s+}\left( X^{\prime }\right)
\right\} \right\rangle ;  \label{A2} \\
F_{\alpha \beta }^{ts}\left( X,X^{\prime }\right) &=&-\varepsilon
^{ts}f_{\alpha \beta }\left( X,X^{\prime }\right) =\left\langle T_{\tau
}\left\{ \psi _{\alpha }^{t}\left( X\right) \psi _{\beta }^{s}\left(
X^{\prime }\right) \right\} \right\rangle ; \\
F_{\alpha \beta }^{+ts}\left( X,X^{\prime }\right) &=&\varepsilon
^{ts}f_{\alpha \beta }^{+}\left( X,X^{\prime }\right) =\left\langle T_{\tau
}\left\{ \psi _{\alpha }^{t+}\left( X\right) \psi _{\beta }^{s+}\left(
X^{\prime }\right) \right\} \right\rangle ,  \notag
\end{eqnarray}%
where $X$ combines space and time and $T_{\tau }$ is the imaginary time
ordered product\cite{AGD}. The time derivatives using the equations of
motion are written via the following commutators:

\begin{eqnarray*}
\frac{\partial G_{\gamma \kappa }^{st}\left( X,X^{\prime }\right) }{\partial
\tau } &=&-\left \langle T_{\tau }\left \{ \left[ H,\psi _{\gamma
}^{s}\left( X\right) \right] \psi _{\kappa }^{t+}\left( X^{\prime }\right)
\right \} \right \rangle -\delta ^{\gamma \kappa }\delta ^{ts}\delta \left(
X-X^{\prime }\right) ; \\
\frac{\partial F_{\gamma \kappa }^{st+}\left( X,X^{\prime }\right) }{%
\partial \tau } &=&\left \langle T_{\tau }\left \{ \left[ H,\psi _{\gamma
}^{s+}\left( X\right) \right] \psi _{\kappa }^{t+}\left( X^{\prime }\right)
\right \} \right \rangle .
\end{eqnarray*}

This leads to

\begin{eqnarray}
\frac{\partial G_{\gamma \kappa }^{st}\left( X,X^{\prime }\right) }{\partial
\tau } &=&-\left\langle T_{\tau }\left\{
\begin{array}{c}
-\int_{r^{\prime \prime }}\ \left\{ iv\sigma _{\gamma \beta }^{i}\delta
_{r_{i}^{\prime \prime }}^{\prime }\left( r-r^{\prime \prime }\right) -\mu
^{\prime }\delta _{\gamma \beta }\delta \left( r-r^{\prime \prime }\right)
\right\} \psi _{\beta }^{s}\left( X^{\prime \prime }\right) \psi _{\kappa
}^{t+}\left( X^{\prime }\right) \\
+\frac{g^{2}}{4}\varepsilon ^{s_{1}s}\varepsilon ^{s_{3}s_{2}}\psi _{\alpha
}^{\dagger s_{1}}\left( X\right) \psi _{\alpha }^{s_{3}}\left( X\right) \psi
_{\gamma }^{s_{2}}\left( X\right) \psi _{\kappa }^{t+}\left( X^{\prime
}\right)%
\end{array}%
\right\} \right\rangle  \label{A3} \\
&&-\delta ^{\gamma \kappa }\delta ^{ts}\delta \left( X-X^{\prime }\right) ;
\\
\frac{\partial F_{\gamma \kappa }^{st+}\left( X,X^{\prime }\right) }{%
\partial \tau } &=&\left\langle T_{\tau }\left\{
\begin{array}{c}
\int_{r^{\prime \prime }}\ \left\{ -iv\nabla _{r^{\prime \prime }}^{i}\delta
\left( r^{\prime \prime }-r\right) \sigma _{\alpha \gamma }^{i}-\mu ^{\prime
}\delta \left( r^{\prime \prime }-r\right) \delta _{\alpha \gamma }\right\}
\psi _{\alpha }^{+s}\left( r^{\prime \prime }\right) \psi _{\kappa
}^{t+}\left( X^{\prime }\right) \\
+\frac{g^{2}}{4}\varepsilon ^{s_{1}s_{2}}\varepsilon ^{s_{3}s}\psi _{\alpha
}^{\dagger s_{1}}\left( \mathbf{r}\right) \psi _{\gamma }^{+s_{2}}\left(
\mathbf{r}\right) \psi _{\alpha }^{s_{3}}\left( \mathbf{r}\right) \psi
_{\kappa }^{t+}\left( X^{\prime }\right)%
\end{array}%
\right\} \right\rangle .  \notag
\end{eqnarray}

Here $\mu ^{\prime }=\mu -wp_{x}.$

The correlators are calculated within the gaussian approximation below.

For the same $X$, $F_{\alpha \beta }^{\ast }=-\left \langle T_{\tau }\psi
_{\beta }^{+}\psi _{\alpha }^{+}\right \rangle \rightarrow F_{\alpha \beta
}^{+}=F_{\beta \alpha }^{\ast }$. In gaussian approximation and using the
spin symmetry one obtains and using definition $\Delta _{\alpha \gamma }=-%
\frac{g^{2}}{4}\varepsilon ^{s_{3}s_{2}}\left \langle \psi _{\alpha
}^{s_{3}}\left( X\right) \psi _{\gamma }^{s_{2}}\left( X\right)
\right
\rangle $. Thus one obtains the first Gorkov equation:

\begin{equation}
-\frac{\partial }{\partial \tau }g_{\gamma \kappa }\left( X,X^{\prime
}\right) -iv\sigma _{\gamma \beta }^{i}\ \nabla _{r}^{i}g_{\beta \kappa
}\left( X,X^{\prime }\right) +\mu ^{\prime }g_{\beta \kappa }\left(
X,X^{\prime }\right) +\Delta _{\alpha \gamma }f_{\alpha \kappa }^{+}\left(
X,X^{\prime }\right) =\delta ^{\gamma \kappa }\delta \left( X-X^{\prime
}\right)  \label{A4}
\end{equation}%
Similarly the second equation takes a form:

\begin{equation}
\frac{\partial }{\partial \tau }f_{\gamma \kappa }^{+}\left( X,X^{\prime
}\right) -iv\sigma _{\alpha \gamma }^{i}\nabla _{r}^{i}f_{\alpha \kappa
}^{+}\left( X,X^{\prime }\right) +\mu ^{\prime }f_{\gamma \kappa }^{+}\left(
X,X^{\prime }\right) -\Delta _{\alpha \gamma }^{\ast }g_{\alpha \kappa
}\left( X,X^{\prime }\right) =0  \label{A5}
\end{equation}

For uniform system, using the Fourier transform, $g_{\gamma \kappa }\left(
X\right) =T\sum \nolimits_{\omega p}\exp \left[ i\left( -\omega \tau
+pr\right) \right] g_{\gamma \kappa }\left( \omega ,p\right) $, with
Matsubara frequencies, $\omega _{n}=2\pi T\left( n+1/2\right) $,
one obtains

\begin{equation}
\left( vp^{i}\sigma _{\gamma \beta }^{i}+\left( i\omega +\mu -wp_{x}\right)
\delta _{\gamma \beta }\right) g_{\beta \kappa }\left( \omega ,p\right)
+\Delta _{\alpha \gamma }f_{\alpha \kappa }^{+}\left( \omega ,p\right)
=\delta ^{\gamma \kappa };  \label{A6}
\end{equation}

\begin{equation}
\left( vp^{i}\sigma _{\beta \gamma }^{i}+\left( -i\omega +\mu -wp_{x}\right)
\delta _{\gamma \beta }\right) f_{\beta \kappa }^{+}\left( \omega ,p\right)
-\Delta _{\alpha \gamma }^{\ast }g_{\alpha \kappa }\left( \omega ,p\right)
=0.  \label{A7}
\end{equation}%
The singlet Ansatz%
\begin{equation}
\Delta _{\alpha \gamma }=\sigma _{\alpha \gamma }^{x}\Delta  \label{A8}
\end{equation}

simplifies the equations:

\begin{eqnarray}
\left( vp^{i}\sigma _{\gamma \beta }^{i}+\left( i\omega +\mu -wp_{x}\right)
\delta _{\gamma \beta }\right) g_{\beta \kappa }\left( \omega ,p\right)
+\Delta \sigma _{\alpha \gamma }^{x}f_{\alpha \kappa }^{+}\left( \omega
,p\right) &=&\delta ^{\gamma \kappa };  \label{A9} \\
\left( vp^{i}\sigma _{\beta \gamma }^{i}+\left( -i\omega +\mu -wp_{x}\right)
\delta _{\gamma \beta }\right) f_{\beta \kappa }^{+}\left( \omega ,p\right)
-\Delta ^{\ast }\sigma _{\gamma \alpha }^{x}g_{\alpha \kappa }\left( \omega
,p\right) &=&0.
\end{eqnarray}%
Casting this in the matrix form,%
\begin{eqnarray}
\left( vp^{i}\sigma ^{i}+i\omega +\mu -wp_{x}\right) \widehat{g}+\Delta
\sigma ^{x}\widehat{f}^{+} &=&I;  \label{n7} \\
\left( vp^{i}\sigma ^{ti}-i\omega +\mu -wp_{x}\right) \widehat{f}^{+}-\Delta
^{\ast }\sigma ^{x}\widehat{g} &=&0,
\end{eqnarray}%
it is easily solved. Substituting $g$ from the second equation,

\begin{equation}
\sigma ^{x}\left( vp^{i}\sigma ^{ti}+\left( -i\omega +\mu -wp_{x}\right)
I\right) \widehat{f}^{+}=\Delta ^{\ast }\widehat{g},  \label{A10}
\end{equation}%
one explicitly obtains the anomalous correlator

\begin{equation}
\widehat{f}^{+}=M^{-1}\Delta ^{\ast }.  \label{A11}
\end{equation}%
where
\begin{equation}
M=\left( vp^{i}\sigma ^{i}+i\omega +\mu -wp_{x}\right) \sigma ^{x}\left(
v\sigma ^{tj}p^{j}+\left( -i\omega +\mu -wp_{x}\right) \right) +\Delta
\Delta ^{\ast }\sigma ^{x}\text{.}  \label{A12}
\end{equation}%
The correlator determining the excitation spectrum consequently is

\begin{equation*}
\widehat{g}=\sigma ^{x}\left( p^{i}\sigma ^{ti}+\left( -i\omega +\mu
-wp_{x}\right) I\right) M^{-1}\Delta ^{\ast }\text{.}
\end{equation*}%
The dispersion relations in the text are defined by zeroes of the
determinant,

\begin{equation}
det\left[ M\right] =-\left( \Delta ^{2}+\left( vp-\mu +wp_{x}\right)
^{2}+\omega ^{2}\right) \left( \Delta ^{2}+\left( vp+\mu -wp_{x}\right)
^{2}+\omega ^{2}\right) \text{.}  \label{A13}
\end{equation}

In this section, the details of the calculation of the superconducting
gap at zero temperature and of the critical temperature are given.

\section{Appendix B. Zero temperature gap.}

At zero temperatures the gap is determined by the self consistency equation:

\begin{equation}
\Delta ^{\ast }=\frac{\Delta ^{\ast }}{2\left( 2\pi \right) ^{3}}\int
d\omega d\mathbf{p}\text{Tr}\left[ \sigma _{x}M^{-1}\right]  \label{B1}
\end{equation}%
At zero temperature the summation over Matsubara frequencies was replaced by
integration. The resulting integral,

\begin{equation}
\frac{1}{g^{2}}=\frac{1}{\left( 2\pi \right) ^{3}}\int d\omega d\mathbf{p}%
\frac{v^{2}p^{2}+\Delta ^{2}+\left( \mu -wp_{x}\right) ^{2}+\omega ^{2}}{%
\left( \Delta ^{2}+\left( vp+wp_{x}-\mu \right) ^{2}+\omega ^{2}\right)
\left( \Delta ^{2}+\left( vp-wp_{x}+\mu \right) ^{2}+\omega ^{2}\right) }%
\text{,}  \label{B2}
\end{equation}%
is first performed over $\omega $:

\begin{equation}
\frac{1}{g^{2}}=\frac{1}{4\left( 2\pi \right) ^{2}}\int_{\theta =0}^{2\pi
}d\theta \int pdp\left\{
\begin{array}{c}
\frac{1}{\left( \Delta ^{2}+\left( \varepsilon -\mu \right) ^{2}\right)
^{1/2}}+\frac{1}{\left( \Delta ^{2}+\left( 2wp_{x}+\varepsilon +\mu \right)
^{2}\right) ^{1/2}} \\
-\frac{1}{\left( \Delta ^{2}+\left( \varepsilon -\mu \right) ^{2}\right)
^{1/2}+\left( \Delta ^{2}+\left( 2wp_{x}+\varepsilon +\mu \right)
^{2}\right) ^{1/2}}%
\end{array}%
\right\}  \label{B3}
\end{equation}%
The resulting expression is already given in polar coordinates and in
addition the energy as function of the polar angle%
\begin{equation}
\varepsilon \left( p,\theta \right) =vp+wp_{x}=vp\left( 1+\kappa \cos \theta
\right) \text{,}  \label{B4}
\end{equation}%
was used. Here $\kappa =\frac{w}{v}.$

In the adiabatic (BCS) limit, when $\mu >>\Omega >>\Delta $, and the
integration is limited to the shell shown in Fig. 4 in both phases. Changing
the integration variable to $\varepsilon $, one obtains:

\begin{eqnarray}
\frac{1}{g^{2}} &=&\frac{1}{4\left( 2\pi \right) ^{2}v^{2}}\int_{0}^{2\pi
}d\theta \frac{sign\left( 1+\kappa \cos \theta \right) }{\left( 1+\kappa
\cos \theta \right) ^{2}}\int_{\varepsilon =-\Omega }^{\Omega }\left(
\varepsilon +\mu \right) \frac{d\varepsilon }{\left( \Delta ^{2}+\varepsilon
^{2}\right) ^{1/2}}=  \label{B5} \\
&=&\frac{\mu }{4\left( 2\pi \right) v^{2}}f\left( \kappa \right)
\int_{\varepsilon =-\Omega }^{\Omega }\frac{d\varepsilon }{\left( \Delta
^{2}+\varepsilon ^{2}\right) ^{1/2}} \\
&=&\ \frac{\mu }{4\pi v^{2}}f\left( \kappa \right) \log \left[ \frac{\Omega +%
\sqrt{\Delta ^{2}+\Omega ^{2}}}{\Delta }\right] \text{,}
\end{eqnarray}%
Here the integration over azimuthal angle $\theta $ was obtained by
replacing variables $\theta $ by $x=v/w+\cos \theta .$ In the case $w/v<1$
(type I semimetal phase), $\ x>0$ and integral over the azimuthal angle reads%
\begin{equation}
f\left( \kappa \right) =\frac{1}{\pi }\int_{\theta }\frac{\ 1}{\left(
1+\kappa \cos \theta \right) ^{2}}=\frac{2}{\left( 1-\kappa ^{2}\right)
^{3/2}}  \label{B6}
\end{equation}%
In the case $w>v$ (type II semimetal phase) the integral might be modified
as
\begin{eqnarray}
f\left( \frac{w}{v}\right) &=&-\frac{1}{\pi }\int_{x=v/w+1}^{v/w-1}dx\frac{%
sign\left( x\right) }{\sqrt{1-\left( x-v/w\right) ^{2}}x^{2}}=  \label{B7} \\
&=&\frac{1}{\pi }\int_{x=v/w-1}^{0}\frac{dx}{\sqrt{1-\left( x-v/w\right) ^{2}%
}x^{2}}-\frac{1}{\pi }\int_{0}^{v/w+1}\frac{dx}{\sqrt{1-\left( x-v/w\right)
^{2}}x^{2}}
\end{eqnarray}%
Represented Eq. (\ref{B7}) in equivalent form

\begin{eqnarray}
&&\lim_{\delta \rightarrow 0}\frac{1}{\pi }\int_{x=\delta }^{-v/w+1}\frac{dx%
}{x^{2}}\left( \frac{1}{\sqrt{1-\left( x+1/w\right) ^{2}}}-\frac{1}{\sqrt{%
1-\left( x-1/w\right) ^{2}}}\right) -  \label{B8} \\
&&-\frac{1}{\pi }\int_{-v/w+1}^{v/w+1}\frac{dx}{\sqrt{1-\left( x-v/w\right)
^{2}}x^{2}}
\end{eqnarray}

and performing the integration in (\ref{B8}), we obtain

\begin{subequations}
\begin{equation}
f\left( \kappa \right) =\frac{2\kappa ^{2}}{\ \pi \left( \kappa
^{2}-1\right) ^{3/2}}\left\{ 2\sqrt{1+\kappa }-1+\log \frac{2\left( \kappa
^{2}-1\right) }{\kappa \left( 1+\sqrt{1+\kappa }\right) ^{2}\delta }\right\}
\label{B9}
\end{equation}

from which the Eq. (18) of the paper follows. Here $\delta =\pi a\Omega
/w\hbar $ is the cutoff parameter.

\section{Appendix C. Critical temperature.}

The critical temperature is defined from the second Gorkov equation
with vanishing $\Delta $:

\end{subequations}
\begin{equation}
\frac{1}{g^{2}}=\frac{T}{\left( 2\pi \right) ^{2}}\int d\mathbf{p}\sum
\nolimits_{n}\frac{v^{2}p^{2}+\left( \mu -wp_{x}\right) ^{2}+\omega _{n}^{2}%
}{\left( \left( vp+wp_{x}-\mu \right) ^{2}+\omega _{n}^{2}\right) \left(
\left( vp-wp_{x}+\mu \right) ^{2}+\omega _{n}^{2}\right) }  \label{CT1}
\end{equation}%
Performing summation over $\omega _{n}$, one obtains,$\ \ $%
\begin{equation}
\frac{1}{g^{2}}=\frac{1}{4\left( 2\pi \right) ^{2}}\int_{\theta =0}^{2\pi
}\int pdp\text{ }\left \{ \frac{\tanh \left[ \frac{\left \vert p\left(
v+w\cos \theta \right) -\mu \right \vert }{2T}\right] }{\left \vert p\left(
v+w\cos \theta \right) -\mu \right \vert }+\frac{\tanh \left[ \frac{\left
\vert p\left( v+w\cos \theta \right) +\mu \right \vert }{2T}\right] }{\left
\vert p\left( v+w\cos \theta \right) +\mu \right \vert }\right \} \text{.}
\label{ct2}
\end{equation}%
In the adiabatic approximation, $\mu >>\Omega $, we get as in the previous
case

$\ $%
\begin{equation}
\frac{1}{g^{2}}=\frac{\mu }{8\pi v^{2}}f\left( \kappa \right) \left\{
2\left( \log \frac{\Omega }{2T_{c}}\text{ }\tanh \left[ \frac{\Omega }{2T_{c}%
}\right] -\int_{\varepsilon =0}^{\Omega }d\varepsilon \frac{\log \varepsilon
}{\cosh ^{2}\left[ \frac{\varepsilon }{2T_{c}}\right] }\right) +\frac{\Omega
}{\mu }\right\} \text{.}  \label{CT3}
\end{equation}%
In this limit Eq.$\left( \text{\ref{CT3}}\right) $ reads for critical
temperature

\begin{equation}
T_{c}=\frac{\pi }{4\gamma _{E}}\Omega \exp \left[ -\frac{4\pi v^{2}}{\ \mu
g^{2}f\left( \kappa \right) }\right] \text{,}  \label{CT4}
\end{equation}%
where $\log \left[ \gamma _{E}\right] =0.577.$

\end{document}